# THE SUSTAINABILITY SOLUTION TO THE FERMI PARADOX

Jacob D. Haqq-Misra[*]
*Department of Meteorology & Astrobiology Research Center*
*The Pennsylvania State University*

Seth D. Baum
*Department of Geography & Rock Ethics Institute*
*The Pennsylvania State University*

No present observations suggest a technologically advanced extraterrestrial intelligence (ETI) has spread through the galaxy. However, under commonplace assumptions about galactic civilization formation and expansion, this absence of observation is highly unlikely. This improbability is the heart of the Fermi Paradox. The Fermi Paradox leads some to conclude that humans have the only advanced civilization in this galaxy, either because civilization formation is very rare or because intelligent civilizations inevitably destroy themselves. In this paper, we argue that this conclusion is premature by introducing the "Sustainability Solution" to the Fermi Paradox, which questions the Paradox's assumption of faster (*e.g.* exponential) civilization growth. Drawing on insights from the sustainability of human civilization on Earth, we propose that faster-growth may not be sustainable on the galactic scale. If this is the case, then there may exist ETI that have not expanded throughout the galaxy or have done so but collapsed. These possibilities have implications for both searches for ETI and for human civilization management.

[*] *Email address*: misra@meteo.psu.edu

# 1. INTRODUCTION

The classic Fermi Paradox can lead to the conclusion that humans have formed the first advanced civilization in the galaxy because extraterrestrial intelligence (ETI) has not yet been observed [1]. Numerous resolutions to this paradox have been proposed [2], spanning the range of cosmological limits to sociological assumptions. A popular class of solutions assumes that the evolution of life is rare in the Universe: Earth may not be wholly unique, but other inhabited planets in the Universe could be too far away for any interaction or detection [3]. But if life is a common phenomenon in the galaxy, then it seems reasonable to expect observable evidence. Furthermore, if the evolution of intelligence is commonplace, then there is hope for projects such as the search for extraterrestrial intelligence (SETI), even though no present observations suggest a technologically advanced ETI has spread throughout the galaxy.

The conclusion that other ETI do not exist contains implicit assumptions about the nature and pattern of ETI. Specifically, this argument requires that ETI expand exponentially from their home location throughout the entire galaxy [4], an assumption that is based on observations of the expansion of human civilization on Earth. The assumption of exponential or other faster-growth is crucial to the conclusion that extraterrestrial civilizations should have colonized the galaxy by now.

However, a closer look at human civilization suggests two problems with this assumption. First, where human populations are exponentially expansive, they often—perhaps always—do so unsustainably, *i.e.* in a way that leads to an eventual end to the exponential expansion. Second, not all human populations are exponentially expansive, such as the !Kung San of the Kalahari Desert [5]. These slower-growth human populations are without question intelligent. Indeed, global human population growth is currently slowing, and humanity as a whole may be transitioning towards a slower-growth, sustainable development pattern. A slower-growth humanity would even remain capable of space colonization.

It is possible that extraterrestrial civilizations face similar sustainability constraints. This possibility suggests a resolution to the Fermi Paradox, which we name the "Sustainability Solution":

> *The "Sustainability Solution" to the Fermi Paradox*: The absence of ETI observation can be explained by the possibility that exponential or other faster-growth is not a sustainable development pattern for intelligent civilizations.

If the Sustainability Solution is true, *i.e.* if intelligent civilizations cannot sustain exponential growth, then no exponentially expansive civilizations should likely be observed. However, the Sustainability Solution does not rule out the possibility of civilizations following slower-growth patterns. Such slower-growth civilizations expand sufficiently slowly that they would not necessarily have colonized the entire galaxy by now. The Sustainability Solution also does not rule out the possibility of faster-growth civilizations colonizing the galaxy and then collapsing. The existence of slower-growth or collapsed civilizations is thus consistent with the lack of human observations of extraterrestrial civilization.

**2. HUMAN CIVILIZATION**

Since its origins in central Africa, humanity has gone through tremendous expansion. It has expanded through space, inhabiting most terrestrial regions of Earth and exploring land, sea, sky, and beyond. It has expanded in population, making it among the most numerous of the large animals [6]. It has expanded in environmental impact, causing some to call the present era of Earth's history the "Anthropocene" in recognition of human civilization now being a dominant force in global environmental change [7]. Finally, it has expanded in resource consumption.

The Fermi Paradox ultimately concerns the spatial expansion of civilizations, but spatial expansion is closely linked with expansion in population, environmental impact, and resource consumption. For example, migration is often driven by resource shortages, which in turn may result from large population and/or environmental degradation. Likewise, migration to uninhabited regions can lead to resource surpluses, which can in turn drive population growth. Finally, broadly expansionist policy can cause expansion in each of space, population, environmental impact, and resource consumption. For a broad and authoritative discussion of these issues, see [8].

In human populations, exponentially expansive practices are commonly considered unsustainable [9, 10]. In order for a development to be sustainable, growth in resource consumption must not exceed growth in resource production—otherwise, resources will eventually be depleted. Resource production on Earth is at most constant: Earth is finite in mass and receives solar radiation at a constant rate. Thus, human civilization cannot indefinitely sustain exponential growth in the consumption of Earth's resources.

The consequences of unsustainable development are often dire. In many documented cases, resource depletion caused by human activities has led to the permanent collapse of human populations [11], and resource depletion and environmental degradation can also cause or exacerbate violent conflict [12]. Note that collapsed human populations do not necessarily disappear—they may persist in diminished numbers. This is particularly evident in the case of Easter Island, where resource depletion is believed to have caused or significantly contributed to a major population decline [11]. Some analysts are concerned that the unsustainable practices of human civilization could lead to a global-scale collapse [10]. Should such a collapse occur, human civilization would not be able to colonize the galaxy.

All hope is not lost for human civilization. There are many documented cases of human populations managing their resources sustainably and achieving long-term survival [13]. On the global scale, human population growth is declining, with a peak of around 9 billion projected for approximately 2075 [14]. Meanwhile, increasing attention is being given to sustainable development [9]. Should human civilization successfully transition to sustainable development, it would have the opportunity to colonize the galaxy.

It is not clear whether unsustainable development will cause a global collapse of human civilization. What is clear is that both sustainable and unsustainable development seem possible. Furthermore, unsustainable development appears closely correlated with exponential growth in space, population, environmental impact, and resource consumption. It is at a minimum plausible that extraterrestrial intelligence would face similar sustainability issues, in which case exponential growth may not be a sustainable development pattern for intelligent civilizations in general.

## 3. EXTRATERRESTRIAL CIVILIZATION

The Fermi Paradox posits that if intelligent life were common in the Universe, then in all likelihood there would exist some extraterrestrial intelligence (ETI) capable of interstellar travel. This ETI would then explore and colonize the galaxy, just as humans have explored and colonized Earth and have begun exploring the Solar System. The magnitude of time required for a technological ETI to spread throughout the galaxy is on the order of 1-100 Myr [4, 15], significantly less than the ~10 Gyr age of the galactic thin disk, so the question arises: *where are they?* If they exist, advanced ETI could have colonized the galaxy several times over by now, so the lack of evidence for their presence implies their non-existence. In syllogistic form, the Fermi Paradox can be expressed following [16] where $A$ = ETI exist, $B$ = ETI are here, and $C$ = ETI are observed:

S1:     If $A$, then (probably $B$)
          If (probably $B$), then (probably $C$)

> Not-(probably *C*)
> Therefore not-(probably *B*)
> Therefore not-*A*

This inference can be criticized because it is only correct if *not-(probably C)* is true. If *(probably C)* is an indeterminate statement, though, then the so-called paradox is logically invalid [16]. For example, ETI exploration of the galaxy could take the form of messenger probes that may have already reached the Solar System, residing in the asteroid belt, Lagrange points, or other stable orbits [17, 18, 19]. Such probes with a limiting size of only ~1-10 meters may have so far eluded observation. If ETI exploration takes such a remote form, then artifacts in the Solar System may yet be observed, but ETI colonization of the Solar System, so far as we know, has not occurred.

Technological ETI are typically assumed to explore and colonize the galaxy just as humans have explored and colonized Earth. This expansion implicitly assumes an exponential growth pattern, leading to the colonization of the entire galaxy:

> Assume that we eventually send expeditions to each of the 100 nearest stars. (These are all within 20 light-years of the Sun.) Each of these colonies has the potential of eventually sending out their own expeditions, and their colonies in turn can colonize, and so forth. If there were no pause between trips, the frontier of space exploration would then lie on the surface of a sphere whose radius was increasing at a speed of 0.10 c. At that rate, most of our Galaxy would be traversed within 650 000 years. [1:133]

The assumption of exponential growth is in turn based on observations of the expansion of human civilization on Earth:

> If, the argument goes, there were intelligent beings elsewhere in our Galaxy, then they would eventually have achieved space travel, and would have explored and colonized the Galaxy, as we have explored and colonized the Earth. [1:128]

However, as discussed above, exponential human population growth and colonization of the planet may not be a sustainable development pattern. This fact calls into question a core justification for the assumption of exponential expansion of ETI civilizations. If ETI civilizations share similar development issues as human civilization, as is assumed in the Fermi Paradox, then ETI civilizations would not be able to sustain exponential expansion [20]. Likewise, if exponential expansion could not be sustained, then ETI civilizations would either have switched to a slower-growth development pattern or collapsed. Collectively, these possibilities suggest the "Sustainability Solution" to the Fermi Paradox: The absence of ETI observation can be explained by the possibility that exponential growth is not a sustainable development pattern for intelligent civilizations.

The Sustainability Solution implies that the existence of slower-growth ETI civilizations cannot be ruled out by the lack of observed ETI because these civilizations would grow too slowly to have reached Earth by now. These civilizations may have always followed a slower-growth development pattern, or they may have started with an exponential or other faster-growth growth pattern only to transition towards slower-growth as faster-growth became unsustainable [21]. Both of these development patterns can be observed in human populations [5], suggesting that both could be possible among ETI civilizations. Furthermore, just as slower-growth human populations (including the global human civilization if it transitions successfully towards sustainable development) are highly intelligent and technologically capable, slower-growth ETI may still be as well. Indeed, slower-growth ETI may even possess space colonization capacity, just without having expanded so rapidly as to colonize the entire galaxy.

The Sustainability Solution also implies that ETI civilizations may have previously followed an exponential or other faster-growth development pattern but eventually collapsed. This collapse could occur at the planetary scale, as is suspected may happen to human civilization [10], at the solar system scale, or even at the galactic scale. If the entire galaxy were once colonized by an ETI civilization, then the colonizing civilization must have collapsed in such a way that no evidence of the colonization has been detected. Evidence of such a *graveyard civilization* may still exist and may eventually be detectable by humans using search efforts different from those already attempted. Furthermore, just as human populations sometimes persist in diminished numbers after undergoing collapse, a collapsed ETI civilization may still exist at a smaller scale.

Having considered the sustainability of ETI civilizations, we can now revisit the Fermi Paradox. If exponential or other faster-growth is unsustainable at the sub-galactic scale, then the supposition by Hart [1] and others that advanced ETI civilization could easily colonize the galaxy is false. Alternatively, this supposition could be true if ETI civilizations that colonize the galaxy eventually collapse, but we are unlikely to observe a galactic colony because faster-growth civilizations collapse quickly relative to astronomical timescales. In principle a civilization could colonize the galaxy through faster-growth and then avoid collapse by transitioning towards sustainable slower-growth; however, the absence of observation of galactic civilization suggests that this has not occurred. In either case, the Fermi Paradox cannot rule out the possibility that slower-growth or post-collapse ETI civilizations currently exist.

The Fermi Paradox syllogism (S1) can be reconstructed, then, with $A'$ = faster-growth ETI civilization exists, $B'$ = faster-growth ETI civilization is here, and $C'$ = faster-growth ETI civilization is observed.

S2:      If $A'$, then $B'$
         If $B'$, then $C'$
            Not- $C'$
         Therefore not-$B'$
         Therefore not-$A'$

This revised inference is still not logically valid because it is impossible to prove that faster-growth ETI civilization has not been observed [16]. After all, there are many explanations for the absence of ETI civilization [2].

A popular class of explanations for this absence of observation involves speculation into the behavior or sociology of ETI. For example, a solution known as the *zoo hypothesis* predicts that ETI civilization has set aside Earth as an undisturbed wildlife preserve [22], stealthily observing Earth (perhaps using a virtual planetarium [23]) and waiting for its inhabitants to cross a technological threshold before making themselves known [24]. A recent hypothesis involving common economic assumptions [25] proposed a solution derived from resource issues, concluding that ETI, like humans, will necessarily lack the patience required to conserve resources for space colonization. Testing such hypotheses may require future technology; for example, the zoo hypothesis might not be falsified (or vindicated) until humans begin interstellar exploration. Nevertheless, most solutions of this class are falsifiable and thus legitimate avenues of scientific inquiry [26].

Other possible explanations invoke the non-linearity of migration. If colonization through the galaxy proceeds as a percolation problem, then expansion should halt after a finite number of colonies [27], resulting in sub-galactic scale clusters around the parent star. Under this scenario, colonized regions of the galaxy would remain isolated from each other, even in a galaxy teeming with intelligent life. Alternatively, a relatively young civilization that engages in economic interstellar travel may find its rapid expansion self-limited by the speed of light [28]. Civilizations that pursue aggressive growth may quickly collapse because growth outpaces migration, while ETI that grow with the limits of the carrying capacity may expand too slow to have colonized the galaxy yet. The persistence hypothesis [29] suggests ETI civilization remains undetected because the solar vicinity is persistently unvisited by ETI civilization—just as regions of Earth such as the Amazon Basin, Siberia, and Indonesian islands are largely untouched by the global human civilization. Persistent sites may remain persistent for a long time, explaining the lack of ETI civilization in the neighborhood of the Sun. Many factors including these may limit the expansion of ETI civilization at the sub-galactic scale. If any ETI civilization overcomes such barriers, then the Sustainability Solution predicts an upper limit to faster-growth galactic expansion.

The classic Fermi Paradox can now be rephrased to account for its implicit assumptions. If faster-growth development is unsustainable, then a faster-growth ETI civilization could expand throughout the galaxy, only to collapse shortly thereafter. As a result, we would likely not observe such a short-lived ETI civilization. This leads us to the inference that *exponentially expansive ETI civilization* does not exist—contrary to the classic conclusion that ETI do not exist at all. However, the non-existence of exponentially expansive ETI civilization does not preclude the existence of ETI. Just as there are human populations maintaining sustainable, slower-growth development, it is entirely possible that ETI exist with slower-growth development patterns. Likewise, just as human populations sometimes persist in diminished numbers after a collapse, it is possible that there exist post-collapse ETI.

**4. IMPLICATIONS FOR SETI**

The Sustainability Solution suggests a recalibration of the human search for ETI, focusing on slower-growth and post-collapse ETI. Each of these forms of ETI would likely yield different signs of their existence, which in turn could be detectable through different strategies.

Traditional SETI projects search for electromagnetic signals broadcast from ETI civilizations [30]. Electromagnetic signals could be broadcast by slower-growth ETI civilizations, just as human civilization would retain the capacity to broadcast signals if it transitions to slower-growth sustainable development. Electromagnetic signals could also be broadcast by post-collapse graveyard civilizations: if part of the population survives the collapse, then the survivors could make *graveyard broadcasts*. Alternatively, if the collapse leaves no survivors, then the signal could, at least in principle, be broadcast by an automatic system deployed before the collapse.

Another approach is to search for terrestrial planets whose atmospheric spectral signatures suggest a higher likelihood of life on the planet [31]. Atmospheric composition alone cannot conclusively demonstrate the presence of life on a distant planet, nor can they necessarily distinguish between intelligent and non-intelligent life, but certain spectral signatures would be unlikely in an abiotic world. For example, the presence of $O_3$ and $O_2$ could be a good biomarker, especially if coupled with atmospheric $CH_4$, and anoxic atmospheres analogous to the early Earth may also be suitable candidates for life [32]. Additionally, the red edge of chlorophyll is a unique biosignature on Earth [33, 34], and inhabited extrasolar planets may exhibit their own distinctive biosignatures. Such signatures would likely occur for slower-growth ETI civilizations because the civilizations' planets necessarily have life on them. Spectral biomarkers may also occur for post-collapse civilizations; if the collapse has survivors, then, as with slower-growth ETI, the survivors' planets necessarily have life on them. Alternatively, if the collapse leaves no survivors, then the planets may still retain a similar biosignature if non-intelligent or non-technological life persists.

A third search strategy allows for the possibility of remote exploration by ETI civilizations. Though colonization of the galaxy may be problematic, slower-growth ETI could conceivably achieve interstellar exploration using small long-lived probes [19]. Remote interstellar exploration by future humans is at least plausible, foreshadowed by the entry of Voyager into the heliosheath at the edge of the Solar System [35], suggesting that slower-growth ETI with sufficient technology could embark on this form of galactic exploration. Searches for ETI probes known *Solar System SETI*, also called SETA (Search for Extraterrestrial Artifacts) or SETV (Search for Extraterrestrial Visitation) [38], has been proposed at visible [39] and radio [40] wavelengths, capable of detecting probes as small as ~10 meters or less. Calls for a Solar System SETI acknowledge that the possibility of remote ETI exploration is at least as likely as interstellar ETI broadcasts, and a survey of the solar vicinity may be more pragmatic than an all-sky search for encoded messages [36, 37].

The Sustainability Solution suggests that Solar System SETI may be the preferred option in searching for technological ETI. Spectral signatures can be detected even if civilization on the planet has not yet developed the capacity to perform electromagnetic broadcast, and a slower-growth civilization may persist for an extended period of time before gaining broadcast capacity. Additionally, spectral signatures can be detected if a post-collapse civilization loses broadcast capacity, and experience with human civilization suggests that collapse is much more likely to cause loss of broadcast capacity than significant change in long-term atmospheric composition. Nevertheless, remote spectral signatures only provide probable biosignatures at best—far from the confirmation of intelligence or technology elsewhere. Solar System SETI, on the other hand, would search for probes of extraterrestrial origin in our stellar vicinity. Artifacts may originate from an extant slower-growth ETI or an extinct galactic empire, but the discovery of either would be near conclusive evidence of extraterrestrial technology.

Ultimately, assumptions about life in the Universe are heavily based on what we observe on Earth. This is because Earth hosts our only known example of life. However, we cannot rule out the possibility that ETI civilization may follow a development pattern sufficiently different that we wouldn't recognize it even if we detected its signal. Therefore, the implications for SETI discussed here cannot be taken as conclusive.

**5. IMPLICATIONS FOR HUMAN CIVILIZATION MANAGEMENT**

If the absence of observed extraterrestrial civilization is due to the Sustainability Solution, *i.e.* due to the unsustainability of exponential growth patterns, then it is tempting to conclude that human civilization needs to transition to sustainable development in order to avoid collapse. Call this proposition Need-SD. (Need-SD does not imply that humans *should* try to avoid the collapse of their civilization, although the present authors would in fact make this the top civilizational priority. For further discussion of the ethics of human extinction, see [41] and [42].) Need-SD is in line with the numerous existing calls for a transition towards sustainable development (*e.g.* [9, 10]), which, if successful, would help human civilization avoid the collapse that may have destroyed exponentially expansive ETI civilizations. Such a transition would also leave open the opportunity for humans to colonize the galaxy.

However, it is premature to conclude that Need-SD is certainly true. One reason is that human civilization may collapse from factors unrelated to sustainable development, such as a large asteroid impact [43]. Another reason is that the limits to exponential growth may lie beyond the planetary scale. For example, perhaps civilization can safely undergo exponential growth until—but only until—it has colonized a solar system [44, 45]. Finally, it cannot be ruled out that sustained exponential growth is possible even if no galactic civilization has done so. Thus, Need-SD does not logically follow from the lack of observation of ETI.

Even if Need-SD is not logically required by the absence of ETI observation, this absence still makes Need-SD more likely. This is because, given the available evidence, we must place a greater-than-zero chance that the absence of ETI observation is due to the unsustainability of exponential civilization growth patterns. Therefore, the arguments presented in this paper serve to strengthen the Need-SD proposition.

## 6. CONCLUSION

The Fermi Paradox cannot logically conclude that humans are the only intelligent civilization in the galaxy. This is due to the Sustainability Solution to the Fermi Paradox presented here: the absence of ETI observation can be explained by the possibility that exponential growth is not a sustainable development pattern for intelligent civilizations. Thus, the Paradox can only conclude that other intelligent civilizations have not sustained exponential growth patterns throughout the galaxy. It is still possible that slower-growth ETI civilizations exist but have not expanded rapidly enough to be easily detectable by the searches humans have yet made. It is also possible that faster-growth ETI civilizations previously expanded throughout the galaxy but could not sustain this state, collapsing in a way that whatever artifacts they might have left have also remained undetected. Both of these growth patterns can be observed in human civilization, suggesting that they may be possible for ETI civilizations as well.

The Sustainability Solution to the Fermi Paradox has practical implications for both the search for extraterrestrial life and human civilization management. In the search for extraterrestrial life, the Sustainability Solution allows that slower-growth ETI civilizations may still transmit radio or other signals. Furthermore, ambitions such as Solar System SETI may eventually discover extraterrestrial messenger probes residing in the asteroid belt and other stellar orbits. For human civilization management, the Sustainability Solution increases the likelihood that human civilization needs to transition towards sustainable development in order to avoid its own collapse.

**ACKNOWLEDGEMENTS**

The authors would like to thank Jim Kasting, Shawn Domagal-Goldman, and two anonymous reviewers for helpful suggestions on earlier versions of this paper.